\documentclass[conference]{IEEEtran}
\IEEEoverridecommandlockouts
\usepackage{cite}
\usepackage{amsmath,amssymb,amsfonts}
\usepackage{algorithmic}
\usepackage{graphicx}
\usepackage{textcomp}
\usepackage{xcolor}
\def\BibTeX{{\rm B\kern-.05em{\sc i\kern-.025em b}\kern-.08em
    T\kern-.1667em\lower.7ex\hbox{E}\kern-.125emX}}

\usepackage{amsthm}
\newtheorem{thm}{Theorem}
\newtheorem{prop}[thm]{Proposition}

\usepackage{import}
\usepackage{amsmath,amssymb}
\usepackage{mathtools}
\mathtoolsset{showonlyrefs,showmanualtags}
\usepackage{cancel}
\usepackage{graphicx}
\usepackage{bbm}
\usepackage[normalem]{ulem}
\usepackage{hyperref}
\usepackage{amsthm}
\usepackage[ruled]{algorithm2e}
\usepackage{multirow,array}

\usepackage{siunitx}

\usepackage{standalone}
\usepackage{pgfplots}
\usepackage{tikz}
\usepackage{subcaption}

\usepackage{import}
\usepackage{amsmath,amssymb}
\usepackage{mathtools}
\usepackage{etoolbox}

\makeatletter
\let\pragma@iinput=\@iinput
\def\@iinput#1{\xdef\@pragmafile{#1}\pragma@iinput{#1}}
\def\@pragmafile{default}
\def\pragmaonce{%
	\csname pragma@\@pragmafile\endcsname
	\global\expandafter\let \csname pragma@\@pragmafile\endcsname =  
}
\makeatother

\DeclareMathOperator*{\argmin}{arg\,min}

\ifdefined\ceil
\else
\DeclarePairedDelimiter\ceil{\lceil}{\rceil}

\fi
\DeclarePairedDelimiterX{\inprd}[2]{\langle}{\rangle}{#1, #2}

\newcommand{\wt}{\text{wt}}

\newcommand{\EV}{{\mathbb{E}}}

\newcommand{\Rel}{\mathbb{R}}

\newcommand{\Hb}{H_b}

\newcommand{\allzeros}{\mathbf{0}}

\newcommand{\defn}{\mathrel{\overset{\makebox[0pt]{\mbox{\normalfont\tiny\sffamily def}}}{=}}}
\newcommand{\lce}{\text{l.c.e}}
\newcommand{\convexhull}{\text{conv}}

\newcommand{\cc}[1]{\mathcal{#1}}

\newcommand{\calC}{\cc{C}}

\newcommand{\calR}{\cc{R}}

\newcommand{\GFt}{\GF{2}}
\newcommand{\GF}[1]{\mathbb{F}_{#1}}

\definecolor{darkgreen}{rgb}{0.0, 0.5, 0.0}

\newcommand{\stkout}[1]{\ifmmode\text{\sout{\ensuremath{#1}}}\else\sout{#1}\fi}

\newcommand{\fig}[1]{{Fig.~\ref{fig:#1}}}

\newcommand{\secn}[1]{{Sec.~\ref{secn:#1}}}

\makeatletter
\newcommand*\ifcounter[1]{%
	\ifcsname c@#1\endcsname
	\expandafter\@firstoftwo
	\else
	\expandafter\@secondoftwo
	\fi
}
\makeatother

\ifcsmacro{theorem}{}{

	\ifcounter{chapter}{
		
	}{
		
	}
	
	\newcounter{example}[section]
	
	\newcounter{problem}[section]
	
}

\def\enumtheoremstart{\begin{enumerate}[noitemsep,label=(\roman*)]}
	\def\enumtheoremend{\end{enumerate}}

\usepackage{transparent}

\newif\ifshowcomments
\newif\ifshowdeleted

\newcommand{\devnull}[1]{}

\newcommand{\comment}[2][]{\ifshowcomments{\printcomment{#1}{#2}}\else\ignorespaces\fi}
\newcommand{\printcomment}[2]{\incolor{#1}{[[\ifx#1\empty\else#1: \fi#2]]}}

\newcommand{\ifequals}[3]{\ifthenelse{\equal{#1}{#2}}{#3}{}}
\newcommand{\case}[2]{#1 #2} 
\newenvironment{switch}[1]{\renewcommand{\case}{\ifequals{#1}}}{}

\definecolor{darkred}{rgb}{0.8, 0.01, 0.1}
\definecolor{darkgreen}{rgb}{0.0, 0.5, 0.0}
\definecolor{darkorange}{rgb}{0.93, 0.35, 0.1}

\newcommand{\incolor}[2]{\ignorespaces
	\begin{switch}{#1}\ignorespaces
		\case{TA}{\color{darkred}}\ignorespaces
		\case{SD}{\color{darkorange}}\ignorespaces
		\case{SK}{\color{darkgreen}}\ignorespaces
		\case{}{\color{red}}\ignorespaces
		#2
	\end{switch}
}

\usepackage{lscape}

\renewcommand{\Hb}{h}

\pragmaonce  

\newcommand{\hX}{\hat{X}}
\newcommand{\hY}{\hat{Y}}
\newcommand{\hZ}{\hat{Z}}
\newcommand{\tZ}{\tilde{Z}}

\newcommand{\enc}{\cc{E}}
\newcommand{\dec}{\cc{D}}

\def\Korner{K\"orner}
\def\combinefunc{\lambda}

\newcommand{\Rate}[1]{R_{\text{#1}}}
\newcommand{\RR}[1]{\calR_{\text{#1}}}

\def\distMeasureForm{d(a, \combinefunc(b,c))=d(a+b,c)}
\newcommand{\matQZ}{U}
\newcommand{\matPC}{V}
\newcommand{\Qfunc}{Q}
\newcommand{\encfunc}{f}
\newcommand{\decfunc}{g}

\begin{document}
\title{
	Two-terminal source coding with \\common sum reconstruction
	\thanks{This work was supported in part by Huawei Technologies
		Canada; and in part by the Natural Science and Engineering Research Council
		(NSERC) of Canada through a Discovery Research Grant.}
}

\author{\IEEEauthorblockN{Tharindu Adikari}
	\IEEEauthorblockA{
		{University of Toronto} \\
		tharindu.adikari@mail.utoronto.ca
	}
	\and
	\IEEEauthorblockN{Stark Draper}
	\IEEEauthorblockA{
		{University of Toronto} \\
		stark.draper@utoronto.ca
	}
}

\maketitle

\begin{abstract}
We present the problem of two-terminal source coding with Common Sum Reconstruction (CSR). 
Consider two terminals, each with access to one of two correlated sources. 
Both terminals want to reconstruct the sum of the two sources under some average distortion constraint, 
and the reconstructions at two terminals must be identical with high probability. 
In this paper, we develop inner and outer bounds to the achievable rate distortion region of the CSR problem for a doubly symmetric binary source. 
We employ existing achievability results for Steinberg's common reconstruction and Wyner-Ziv's source coding with side information problems, and an achievability result for the lossy version of \Korner-Marton's modulo-two sum computation problem. 
\end{abstract}

\begin{IEEEkeywords}
source coding, common reconstruction, modulo-two sum, lossy \Korner-Marton
\end{IEEEkeywords}


\section{Introduction} \label{secn:twoterminalsourcecoding}

\begin{figure}
	\centering\includegraphics[width=0.8\columnwidth]{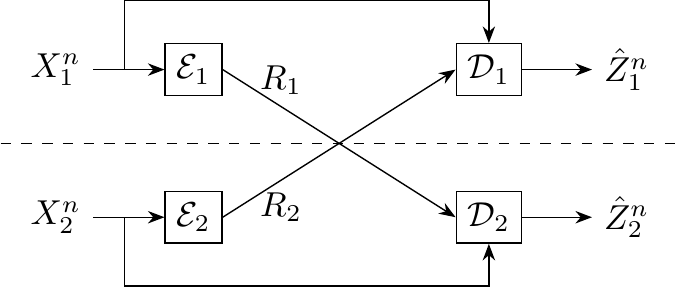}
	\caption{
		The dashed line separates the two terminals. 
		The decoder outputs $\hZ_1^n$ and $\hZ_2^n$, two reconstructions of $Z^n=X_1^n+X_2^n$. 
		The CSR problem is obtained by adding the constraint that $\hZ_1^n=\hZ_2^n$ with high probability. 
	}
	\label{fig:two_terminal_coding}
	\vspace*{-0.063in}
\end{figure}

Consider the two-terminal source coding setup in \fig{two_terminal_coding}, where the terminals operate the encoder-decoder pairs $(\enc_1,\dec_1)$ and $(\enc_2,\dec_2)$. 
Let $\{X_{1,i},X_{2,i}\}_{i=1}^{\infty}$ be a sequence of independent and identically distributed (i.i.d.) pairs of random variables, and let $(X_1, X_2)$ have the same joint distribution as $(X_{1,i},X_{2,i})$. 
We denote by $X_1^n$ and $X_2^n$ the first $n$ observations of $X_{1,i}$ and $X_{2,i}$. 
The sequences $X_1^n$ and $X_2^n$ are observed separately at the two terminals (e.g., both/only $\enc_1$ and $\dec_1$ observe $X_1^n$). 
The two encoders respectively use rates $R_1$ and $R_2$ to encode $X_1^n$ and $X_2^n$. 
The decoders produce $\hZ_1^n$ and $\hZ_2^n$, two reconstructions of the sum $Z^n=X_1^n+X_2^n$, which satisfy $\EV[d(Z^n,\hZ_1^n)]\leq D$ and $\EV[d(Z^n,\hZ_2^n)]\leq D$ for some distortion measure $d(\cdot,\cdot)$ and $D\geq0$. 
We obtain the ``Two-terminal Source Coding with Common Sum Reconstruction'' (CSR) problem by adding the constraint that with high probability $\hZ_1^n$ and $\hZ_2^n$ must be equal, i.e., both decoders must produce a Common Reconstruction (CR) of the sum $Z^n=X_1^n+X_2^n$. 
The goal of this paper is presentation of inner and outer bounds to the achievable rates distortion region (to be defined formally later) in the CSR problem, consisting of achievable rates distortion triples $(R_1,R_2,D)$, when $(X_{1},X_{2})$ is a Doubly Symmetric Binary Source (DSBS) and $d(\cdot,\cdot)$ is Hamming distortion measure.
The summation in this case is modulo-two. 

One may make the following two important observations about the CSR problem. 
First, since $X_{1}$ and $X_{2}$ are correlated, the decoders may use $X_1^n$ and $X_2^n$ as side information at $\dec_1$ and $\dec_2$ respectively. 
Due to the availability of decoder side information, the encoders can lower the required rates $R_1$ and $R_2$. 
Second, decoders only need to reconstruct the sum $X_1^n+X_2^n$. The recovery of $X_1^n$ (at $\dec_2$) and $X_2^n$ (at $\dec_1$) is not needed. 
Reconstructing the sum should not take a larger bit-rate than the rate required for reconstructing the individual sources.

The CSR problem is motivated by its application to algorithms that require computation of the average of correlated random variables. Examples include distributed counterparts of Stochastic Gradient Descent (SGD), power iteration and Lloyd's ($k$-means) algorithm \cite{suresh2017distributed}. 
In practice, algorithms such as butterfly all-reduce \cite{thakur2005optimization} facilitate the computation of averages over a large number of compute nodes. 
For example, let the two terminals in \fig{two_terminal_coding} be two compute nodes optimizing some function with synchronous SGD, and let $X_1$ and $X_2$ be two (scalar) stochastic gradients of the function computed by the $i$th node. 
Two stochastic gradients are correlated since they are noisy estimates of the gradient of the function. 
In synchronous SGD, each compute node wants to update its optimizing variable using the average gradient $(X_1+X_2)/2$. 
\fig{two_terminal_coding} presents a suitable communication setup for this two-terminal setting. 
The butterfly all-reduce algorithm employs the two-terminal communication setup as a basic building block, and uses it recursively to extend the sum computation to an arbitrary number of compute nodes. 
A similar mechanism can be employed for distributed counterparts of power iteration and Lloyd's algorithm as well. Both algorithms require mean computation, and depending on the underlying dataset, one may observe correlations in the vectors that participate in the mean computation.

The remainder of the paper is organized as follows. 
In \secn{Definitionsandnotations} we formally define the CSR problem. 
In \secn{innerboundsCSR}, we develop an inner bound to the achievable CSR rate distortion region by considering existing achievability results for Steinberg's CR problem \cite{steinberg2009coding}. 
In \secn{achievabilityLKM} we develop a new achievability result for the ``lossy'' version of the \Korner-Marton modulo-two sum problem \cite{korner1979encode}, and use this new achievability result to obtain another inner bound. 
In \secn{outerboundsCSR} we develop an outer bound by considering Wyner-Ziv's problem of source coding with side information \cite{wyner1976rate}. 
In \secn{comparebounds} we compare the inner bounds and the outer bound for different DSBS sources. 
Finally, in \secn{conclusion} we present concluding remarks. 

\section{Definitions and notations} \label{secn:Definitionsandnotations}

Since in this paper we focus on binary sources, we present the formal definition of the CSR problem for binary sources as follows. For a positive integer $N$, let $[N]$ denote the index set $\{1,\dots,N\}$, and let $\GFt$ denote the finite field of order $2$. 
An $(n,N_1,N_2,D,\epsilon)$-CSR coding scheme consists of 
two encoding functions 
$\enc_1:\GFt^n\to[N_1]$ and 
$\enc_2:\GFt^n\to[N_2]$, 
and two decoding functions 
$\dec_1:\GFt^n\times[N_2]\to\GFt^n$ and 
$\dec_2:\GFt^n\times[N_1]\to\GFt^n$, 
such that for 
$\hZ_1^n = \dec_1(X_1^n, \enc_2(X_2^n))$, 
$\hZ_2^n = \dec_2(X_2^n, \enc_1(X_1^n))$ and 
$Z^n=X_1^n+X_2^n$, 
we have 
$\EV[d(Z^n,\hZ_1^n)]\leq D$ and 
$\EV[d(Z^n,\hZ_2^n)]\leq D$, 
and $\Pr(\hZ_1^n\neq\hZ_2^n)\leq\epsilon$, 
where $d(\cdot,\cdot)$ is the Hamming distortion measure. 
A specific CSR coding scheme is denoted by the four mappings $(\enc_1, \enc_2, \dec_1, \dec_2)$. 

A scalar triple $(R_1, R_2, D)$ is called CSR-achievable if for every $\epsilon>0$ and $n$ 
sufficiently large there exists an $(n,2^{n(R_1+\epsilon)},2^{n(R_2+\epsilon)},D,\epsilon)$-CSR coding scheme. 
We denote by $\RR{CSR}\subset\Rel^3$ the set of all CSR-achievable triples. 

A source $(X,Y)$ is a DSBS$(p)$ if $X$ and $Y$ are binary sources that satisfy 
$\Pr(X=0,Y=0)=\Pr(X=1,Y=1)=(1-p)/2$ and $\Pr(X=0,Y=1)=\Pr(X=1,Y=0)=p/2$ 
for some $0\leq p\leq\frac{1}{2}$. 
We denote the binary entropy function by $\Hb(\cdot)$, i.e., $\Hb(p)=-p\log_2 p - (1-p)\log_2 (1-p)$, 
and the binary convolution operator by $*$, i.e., $p*q = p(1-q)+(1-p)q$.

\section{Inner bound with Steinberg's CR problem} \label{secn:innerboundsCSR}

\begin{figure}
	\centering\includegraphics[width=0.68\columnwidth]{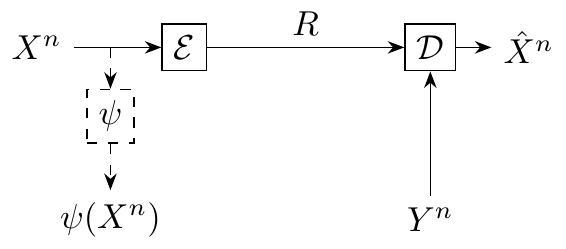}\\~\\
	\caption{
		Steinberg's CR problem.
	}
	\label{fig:rate_dist_side_info}
\end{figure}

A set $\RR{}\subset\Rel^3$ provides an inner bound to $\RR{CSR}$ if $\RR{}\subseteq\RR{CSR}$. 
An inner bound to $\RR{CSR}$ can easily be obtained by considering Steinberg's CR problem \cite{steinberg2009coding} presented in \fig{rate_dist_side_info}. 
Steinberg's problem adds one constraint to the problem of source coding with side information considered by Wyner and Ziv in \cite{wyner1976rate}. The constraint is that with high probability the encoder is able to reconstruct the decoder output. 
In \fig{rate_dist_side_info}, the function triple $(\enc,\dec,\psi)$ constitutes a CR code, where 
$\enc$ encodes $X^n$ with rate $R$, 
and $\dec$ reconstructs $X^n$ as $\hX^n$ with the help of side information $Y^n$, 
such that $\EV[d(X^n,\hX^n)]\leq D$ and $\psi(X^n)=\hX^n$ with high probability. 
The last constraint 
limits the decoder's ability to exploit fully the side information $Y^n$, since $Y^n$ is not available at the encoder. 
Other variants of Steinberg's problem have also been considered. 
For example, in \cite{lapidoth2014constrained} the authors generalize Steinberg's CR problem to allow the encoder and decoder reconstructions to be different within some fidelity constraint, 
and in \cite{lu2016binary} the authors consider an encoder that observes side information that is different from the side information available to the decoder.

\begin{figure}
	\centering
	\includegraphics[width=0.9\columnwidth]{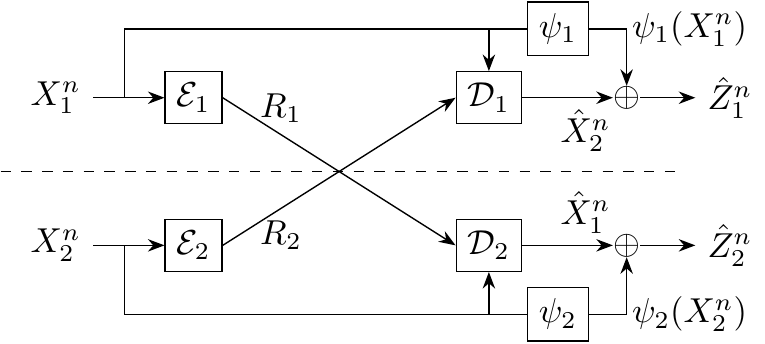}
	\caption{Application of Steinberg's CR problem to CSR.}
	\label{fig:two_steinberg_systems}
\end{figure}

For a DSBS$(p)$ source $(X,Y)$, the rate distortion function for Steinberg's problem in \fig{rate_dist_side_info} is given by $\Rate{CR}(D)$ where $\Rate{CR}(D) = \Hb(p*D) - \Hb(D)$ for $0\leq D\leq\frac{1}{2}$ and $\Rate{CR}(D) = 0$ for $\frac{1}{2}\leq D$. 
We refer the reader to \cite{steinberg2009coding} for more details. 
An inner bound to $\RR{CSR}$ can be obtained by considering two CR codes as we summarize next. 
\begin{prop} \label{prop:propositionSteinberg}
	Let $(X_{1},X_{2})$ be a DSBS$(p)$ with $0\leq p\leq\frac{1}{2}$, 
	\begin{align}
	\RR{A} &= \left\{ (R_1,R_2,D)\ \middle\vert \begin{array}{l}
	0\leq D_1,D_2\leq\frac{1}{2}, \\
	R_1\geq\Rate{CR}(D_1), \\ 
	R_2\geq\Rate{CR}(D_2), \\ 
	D\geq D_1*D_2
	\end{array}\right\}, \text{ and}\\
	\RR{B} &= \{ (R_1,R_2,D) \mid R_1\geq0, R_2\geq0, D\geq p \}.
	\end{align}
	Then, the convex hull $\convexhull(\RR{A}\cup\RR{B})$, 
	is a subset of $\RR{CSR}$ 
	and provides an inner bound to $\RR{CSR}$. 
\end{prop}
The proof of Proposition~\ref{prop:propositionSteinberg} can be sketched as follows. 
First, note that the triples in $\RR{B}$ are CSR-achievable, since for $D\geq p$, the decoders can deterministically set $\hZ_1^n=\hZ_2^n=0$, requiring zero rates for $R_1$ and $R_2$. 
Second, we show later that all triples in $\RR{A}$ are also CSR-achievable. 
Then, any triple in $\convexhull(\RR{A}\cup\RR{B})$ is CSR-achievable by time-sharing between triples in $\RR{A}$ and $\RR{B}$. 
To see that all triples in $\RR{A}$ are CSR-achievable, let $(\enc_1,\dec_2,\psi_1)$ and $(\enc_2,\dec_1,\psi_2)$ be two independent CR codes 
such that $\EV[d(X_1^n,\hX_1^n)]\leq D_1$ and $\EV[d(X_2^n,\hX_2^n)]\leq D_2$. 
Let the two codes be applied to the CSR setup as shown in \fig{two_steinberg_systems}. 
Due to the use of the two CR codes we have $\hZ_1^n=\hZ_2^n$ with high probability. 
Since the two CR codes are selected independently, and due to the Markov chain $\hX_1\--(X_1,X_2)\--\hX_2$, 
the quantization errors $X_1+\hX_1$ and $X_2+\hX_2$ are independent of each other, 
and $\EV[X_1+X_2 + \hX_1 + \hX_2]\leq D_1*D_2$. 
Therefore, 
$\EV[d(Z^n,\hZ_1^n)] \leq D_1*D_2 + \delta$
\comment{
	Letting $\wt(\cdot)$ denote the Hamming weight of the input and by using the triangle inequality we have 
	\begin{align}
	n\EV[d(Z^n,\hZ_1^n)] 
	&= \EV[\wt(X_1^n+X_2^n + \psi(X_1^n) + \hX_2^n)] \\
	&= \EV[\wt(X_1^n+\psi_1(X_1^n)) + \wt(X_2^n+\hX_2^n)] \\
	&\leq nD_1+nD_2 + n\delta, 
	\end{align}}
where $\delta$ is due to the event that $\psi_1(X_1^n)\neq\hX_1^n$, 
and $\delta$ can be made arbitrarily small with the choice of $(\enc_1,\dec_2,\psi_1)$. 
A similar sequence of arguments apply to $\EV[d(Z^n,\hZ_2^n)]$. 
This completes the sketch of the proof of Proposition~\ref{prop:propositionSteinberg}. 
Note that the decoders reconstruct individual sources, and the decoders do not exploit the fact that only the sum of the sources has to be reconstructed. 
Therefore, $\RR{A}$ and $\convexhull(\RR{A}\cup\RR{B})$ provide only inner bounds to $\RR{CSR}$.

\section{Inner bound with Lossy \Korner-Marton (LKM) problem} \label{secn:achievabilityLKM}

\begin{figure}
	\centering\includegraphics[width=0.655\columnwidth]{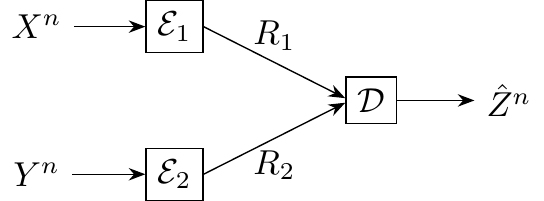}\\~\\
	\caption{Lossy \Korner-Marton problem. The decoder reconstructs $Z^n=X^n+Y^n$ as $\hZ^n$ under some fidelity criterion.}
	\label{fig:korner_marton}
\end{figure}

Another inner bound to $\RR{CSR}$ can be obtained by considering the Lossy \Korner-Marton (LKM) problem presented in \fig{korner_marton}. 
In the LKM problem, the decoder wishes to reconstruct the sum $Z^n=X^n+Y^n$ as $\hZ^n$ under some fidelity criterion. 
The LKM problem is a generalization of the (lossless) \Korner-Marton problem~\cite{korner1979encode}. 
When $(X,Y)$ is a DSBS$(p)$ and $d(\cdot,\cdot)$ is Hamming distance, the authors of \cite{korner1979encode} exploit the linearity of the sum modulo-two computation $Z^n=X^n+Y^n$ and demonstrate that with high probability the decoder can reconstruct $Z^n$ (losslessly) if and only if $R_1\geq\Hb(p)$ and $R_2\geq\Hb(p)$. 
In more recent work, the authors of \cite{sefidgaran2015korner} develop inner and outer bounds to the set of achievable rate pairs $(R_1,R_2)$ in the lossless case when $(X,Y)$ is an asymmetric binary source. The authors of \cite{nair2020optimal} show that the result in \cite{korner1979encode} applies to larger class of binary distributions.

\begin{figure}
	\centering\includegraphics[width=0.845\columnwidth]{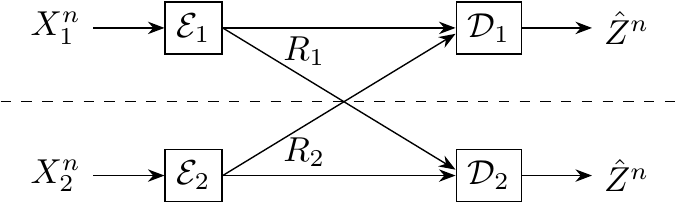}\\~\\
	\caption{Application of LKM problem to CSR.}
	\label{fig:two_terminal_korner_marton}
\end{figure}

The LKM problem can be formally defined as follows. 
An $(n,N_1,N_2,D)$-LKM coding scheme consists of 
two encoding functions $\enc_1:\GFt^n\to[N_1]$ and $\enc_2:\GFt^n\to[N_2]$, 
and a decoding function $\dec:[N_1]\times[N_2]\to\GFt^n$, 
such that for $Z^n=X^n+Y^n$ and $\hZ^n = \dec(\enc_1(X^n), \enc_2(Y^n))$ 
we have $\EV[d(Z^n,\hZ^n)]\leq D$, where $d(\cdot,\cdot)$ is Hamming distortion. 
A specific LKM coding scheme is denoted by the three mappings $(\enc_1, \enc_2, \dec)$. 
A scalar triple $(R_1, R_2, D)$ is called LKM-achievable if for every $\epsilon>0$ and sufficiently large $n$ there exists an $(n,2^{n(R_1+\epsilon)},2^{n(R_2+\epsilon)},D)$-LKM coding scheme. 
We denote by $\RR{LKM}\subset\Rel^3$ the set of all LKM-achievable triples. 

\fig{two_terminal_korner_marton} shows how the LKM coding scheme can be applied to obtain an inner bound to the CSR problem. 
The two decoders in \fig{two_terminal_korner_marton} are identical and they are located at the two terminals. 
With \fig{two_terminal_korner_marton} we observe that every triple in $\RR{LKM}$ is CSR-achievable, i.e., $\RR{LKM}\subseteq\RR{CSR}$. 
Note that $\RR{LKM}$ provides only an inner bound to $\RR{CSR}$ since the decoders in \fig{two_terminal_korner_marton} do not exploit the fact that at least one of the sources is available to each decoder with full fidelity. 
For example, in CSR $\dec_1$ has access to $X_1^n$ (cf. \fig{two_terminal_coding}), 
although in \fig{two_terminal_korner_marton} $\dec_1$ observes $X_1^n$ only with a rate $R_1$. 

To the best of our knowledge, a tight bound for $\RR{LKM}$, i.e., \emph{lossy} \Korner-Marton problem, has not yet been proposed. 
In \cite{krithivasan2009lattices, wagner2010distributed, yang2014distributed} the authors obtain inner bounds to $\RR{LKM}$ for Gaussian sources under squared distortion, and the bounds have been shown to be tight for some special cases. 
The authors of \cite{lim2019towards} present an inner bound for a generalized version of the LKM problem setup that includes an arbitrary number of linear combinations of the sources. The authors employ joint typicality-based encoding and decoding functions to obtain the inner bound. In contrast, in this paper we employ minimum distance-based encoding and decoding functions. The authors of \cite{lim2019towards} do not evaluate the inner bound for the special case of the binary-Hamming LKM problem.

We consider the LKM problem for the special case of binary sources. 
In \secn{proposedcoding} we propose an LKM coding scheme, 
and in \secn{achievabilityofproposed} we characterize the achievable rate distortion region in the proposed scheme, assuming $(X,Y)$ is a DSBS$(p)$. As a corollary we obtain an inner bound to $\RR{CSR}$.

\subsection{Proposed coding scheme} \label{secn:proposedcoding}
The proposed scheme is defined by two matrices $\matQZ\in\GFt^{m\times n}$ and $\matPC\in\GFt^{r\times n}$. 
For a column vector $z^n\in\GFt^n$, let $\matQZ z^n$ and $\matPC z^n$ be the products 
of the matrices with $z^n$ in $\GFt$. 
Let 
$\calC=\{ z^n\in\GFt^n \mid \matQZ z^n=\allzeros \},$ 
where 
$\allzeros$ is the all-zeros column vector of length-$m$, and let 
$\Qfunc(x^n) = \argmin_{z^n\in\calC} \wt(x^n+z^n),$ 
where $x^n+z^n$ is the element-wise sum in $\GFt$ and 
$\wt(\cdot)$ is the Hamming weight of the input. 
Since $\Qfunc(\cdot)$ searches for the element in $\calC$ closest to the input, 
we refer to $\Qfunc(\cdot)$ as the quantizing function 
and to $\calC$ as the quantizing codebook. 
Let 
$\encfunc:\GFt^n\to\GFt^r$ and 
$\decfunc:\GFt^r\times\GFt^r\times\GFt^m\to\GFt^n$ be defined as 
$\encfunc(x^n) = 
\matPC\Qfunc(x^n)$, 
i.e., composition of $\matPC$ with $\Qfunc$, 
and 
\begin{equation} \label{eqn:decoderfunc}
\begin{aligned}
\decfunc(x^r, y^r,s^m) =
\argmin_{z^n\in\GFt^n} \quad &
\wt(z^n) \\
\textrm{s.t.} 
\quad\quad & \matQZ z^n=s^m, ~\matPC z^n = x^r+y^r.
\end{aligned}
\end{equation}
In the proposed scheme, the function $\encfunc$ is used by the encoders and $\decfunc$ by the decoder. 
In comparison to the encoder in \cite{korner1979encode}, $\encfunc$ consists of an additional quantization step that employs the linear codebook $\calC$. 
The linearity of the codebook dovetails with the linear structure of the sum computation, thereby making a rate reduction possible, as we will see in \secn{achievabilityofproposed}. 

\subsection{Achievability of the proposed coding scheme} \label{secn:achievabilityofproposed}
The proposed scheme can be used to show the achievability of the rate region summarized in Proposition~\ref{prop:propositionLKM}. 
\begin{prop} \label{prop:propositionLKM}
	Let $(X, Y)$ be a DSBS$(p)$ with $0\leq p\leq\frac{1}{2}$, and 
	$
	\RR{C} = \{ (R_1,R_2,D) \mid  R_1, R_2\geq \Hb(p*q*q)-\Hb(q), D\geq q*q, 0\leq q\leq \frac{1}{2} \}. 
	$
	For every scalar triple $(R_1, R_2, D)\in\RR{C}$, there exist a pair of matrices $(\matQZ, \matPC)$ (as defined in \secn{proposedcoding}) such that the triple is LKM-achievable with the coding scheme $(\enc_1, \enc_2, \dec)$ in \secn{proposedcoding}. 
	
	Furthermore, the convex hull $\convexhull(\RR{C}\cup\RR{B})$, where $\RR{B}$ is as defined in Proposition~\ref{prop:propositionSteinberg}, is a subset of $\RR{CSR}$ and provides an inner bound to $\RR{CSR}$. 
\end{prop}

The proof of Proposition~\ref{prop:propositionLKM} can be sketched as follows. 
Let $\matQZ$ be the ${m\times n}$ parity check matrix of a `good' linear rate distortion code for a uniform source that incurs average distortion $q$. 
Such a source code will have $m\approx n\Hb(q)$ \cite[Theorem 4]{goblick1963coding, berger1975advances}.  
Let $D_X^n\in\GFt^n$ and $D_Y^n\in\GFt^n$ be two length-$n$ dither sequences sampled independently from the i.i.d. Bernoulli($\frac{1}{2}$) distribution. 
Both dither sequences are also known to the decoder. 
The encoders respectively compute, and send to the decoder, $\encfunc(X^n+D_X^n)$ and $\encfunc(Y^n+D_Y^n)$. 
Let $\hX^n=\Qfunc(X^n+D_X^n)$ and $\hY^n=\Qfunc(Y^n+D_Y^n)$. 
The encoders send $\matPC\hX^n$ and $\matPC\hY^n$ to the decoder, each using $r$ bits (recall $\matPC$ is an ${r\times n}$ matrix). 
Let $W_X^n$ and $W_Y^n$ be the quantization errors of the two sequences, i.e., $W_X^n=\hX^n+X^n+D_X^n$ and $W_Y^n=\hY^n+Y^n+D_Y^n$. 
Adding dither prior to quantization makes $W_X^n$ and $W_Y^n$ independent. 
To see that $W_X^n$ and $W_Y^n$ are not always independent without dither, consider the limiting case $p=0$. 
In this case $X^n=Y^n$ and $W_X^n=\Qfunc(X^n)+X^n=\Qfunc(Y^n)+Y^n=W_Y^n$ with probability $1$. 
Adding dither makes the two quantizer inputs $X^n+D_X^n$ and $Y^n+D_Y^n$ independent, 
which is important to making $W_X^n$ and $W_Y^n$ independent and obtaining the rate expression in Proposition~\ref{prop:propositionLKM}. 

The decoder knows $\matPC\hX^n$, $\matPC\hY^n$, $D_X^n$, $D_Y^n$ and computes 
$\matPC\hX^n+\matPC\hY^n+\matPC D_X^n+\matPC D_Y^n = 
\matPC\hZ^n$, where $\hZ=\hX^n+\hY^n+D_X^n+D_Y^n$. 
Observing that we can reexpress $\hZ$ as $\hZ=X^n+W_X^n+Y^n+W_Y^n=Z^n+W_X^n+W_Y^n$, 
and by observing that $Z^n,W_X^n,W_Y^n$ are jointly independent, 
the marginal statistics of $\hZ$ are Bernoulli($p*q*q$). 
Similarly, $\EV[d(Z,\hZ)]\leq q*q$. 
Per Lemma in \cite{korner1979encode}, there exists a matrix $\matPC$ such that $\hZ$ can be recovered from $\matPC\hZ^n$ if $r>\Hb(p*q*q)$. 
However, this rate can be reduced if the decoder employs $D_X^n$ and $D_Y^n$ as side information. 
Specifically, let $\matQZ\hZ^n = \matQZ\hX^n+\matQZ\hY^n+\matQZ D_X^n+\matQZ D_Y^n=\matQZ(D_X^n+D_Y^n)\defn S^m$, 
where $S^m\in\GFt^m$. 
The decoder computes $\tZ=\decfunc(\matPC\hX^n, \matPC\hY^n, S^m)$. 
Since there exist only around $2^{n(1-\Hb(q))}$ length-$n$ sequences that satisfy $\matQZ\hZ^n=S^m$, 
we only require that $r>\Hb(p*q*q)-\Hb(q)$ to obtain $\tZ=\hZ$ with high probability. 
Therefore, we conclude that all triples in $\RR{C}$ are LKM-achievable, i.e., $\RR{C}\subseteq\RR{LKM}$. 

By following the time-sharing argument as in Proposition~\ref{prop:propositionSteinberg}, we have that all triples in $\convexhull(\RR{C}\cup\RR{B})$ are CSR-achievable, showing that $\convexhull(\RR{C}\cup\RR{B})\subseteq\RR{CSR}$. 
A similar time-sharing argument is used in \cite{wyner1976rate} when obtaining the rate distortion function of a binary source with decoder side information.

\section{Outer bound with Wyner-Ziv problem} \label{secn:outerboundsCSR}

A set $\RR{}\subset\Rel^3$ provides an outer bound to $\RR{CSR}$ if $\RR{CSR}\subseteq\RR{}$. 
An outer bound to $\RR{CSR}$ can be obtained by waiving the CR constraint in the CSR problem. 
Specifically, reconsider the problem setup in \fig{two_terminal_coding} without the CR constraint. 
We refer to this new problem as the ``Two-terminal Sum Estimation'' (TSE) problem. 
Let the achievable rate distortion region for TSE problem be $\RR{TSE}\subset\Rel^3$, with $\RR{TSE}$ consisting of the triples $(R_1,R_2,D)$ that produce the decoder outputs that satisfy $\EV[d(Z^n,\hZ_1^n)]\leq D$ and $\EV[d(Z^n,\hZ_2^n)]\leq D$. 
Due to the waiving of the CR constraint we have that $\RR{CSR}\subseteq\RR{TSE}$. 

When $X_1$ and $X_2$ take values in $\GFt$ and $d(\cdot,\cdot)$ is Hamming distance, one can characterize $\RR{TSE}$ by considering two parallel source coding with side information problems, i.e., two Wyner-Ziv (WZ) systems \cite{wyner1976rate}. 
With reference to \fig{two_terminal_coding}, the first WZ system is from $\enc_1$ to $\dec_2$ and the second WZ system is from $\enc_2$ to $\dec_1$. 
In the first system, $\enc_1$ encodes $X_1^n$ and sends to $\dec_2$ with a rate $R_1$. 
Decoder $\dec_2$ uses $X_2^n$ as side information to reconstruct $X_1^n$ as $\hX_1^n$, and $\dec_2$ outputs $\hZ_2^n=\hX_1^n+X_2^n$. 
The second WZ system works in a similar manner with $\enc_2$ using a rate $R_2$ to send to $\dec_1$, and $\dec_1$ outputting $\hZ_1^n=X_1^n+\hX_2^n$. 

Let $g(D) = \Hb(p*D) - \Hb(D)$ if $0\leq D<p$ and $g(D) = 0$ if $p\leq D$. 
Let $\Rate{WZ}(D)$ be 
the minimum rate required by the WZ scheme for decoder $\dec_j$, $j\in\{1,2\}$, to produce $\hZ_j$, satisfying $\EV[d(Z,\hZ_j)]\leq D$. 
In \cite{wyner1976rate}, the authors show that $\Rate{WZ}(D) = \lce(g(D))$, 
the lower convex envelope of $g(D)$ (see Sec.~II-A \cite{wyner1976rate} and Example~1 in \cite{steinberg2009coding}). 

Note that when using the WZ scheme the decoders do not exploit the fact that only the sum of the sources, $Z=X_1+X_2$, needs to be reconstructed. 
Although, this may make one think that the WZ scheme does not provide a tight bound for $\RR{TSE}$. This is not the case as we will see next. 
For example, considering $\dec_1$, it can be shown that there exists no scheme which with it is possible for $\dec_1$ to obtain an reconstruction of $Z$, say $\tilde{Z}_1\in\GFt$, with a rate $\Rate{WZ}(D)$, such that $\EV[d(Z,\tilde{Z}_1)]\leq D'<D$. 
To see this, assume there exists such a scheme. 
Then, 
$d(X_2,(X_1+\tilde{Z}_1))
= d(X_2+X_1+\tilde{Z}_1) 
= d(Z,\tilde{Z}_1)
$,
which yields
$\EV[d(X_2,(X_1+\tilde{Z}_1))]\leq D'<D$. 
This result contradicts the WZ converse, since such a result would imply that $(X_1+\tilde{Z}_1)$, which can be computed at $\dec_1$, is a better reconstruction of $X_2$ that can be obtained with rate $\Rate{WZ}(D)$. 
A similar sequence of arguments is applicable to the reconstruction of $\hZ_2$ at $\dec_2$ as well. 
Therefore, the two WZ systems can be used to obtain a tight characterization of $\RR{TSE}$ when $X_1$ and $X_2$ are in $\GFt$, and $d(\cdot,\cdot)$ is Hamming distance. 

\begin{figure}[h]
	\centering\includegraphics[width=.48\textwidth]{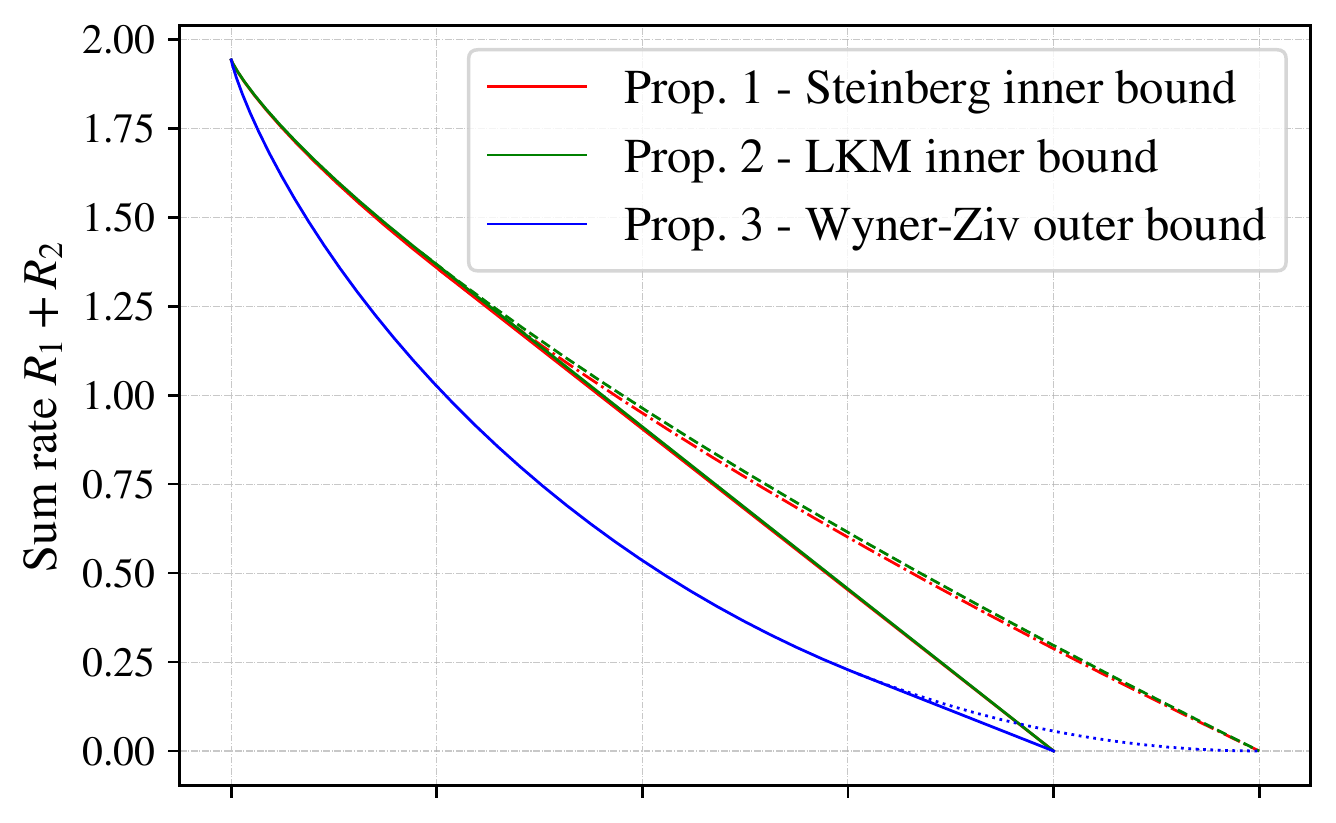}
	~\\ \vspace{-0.7em}
	\centering\includegraphics[width=.473\textwidth]{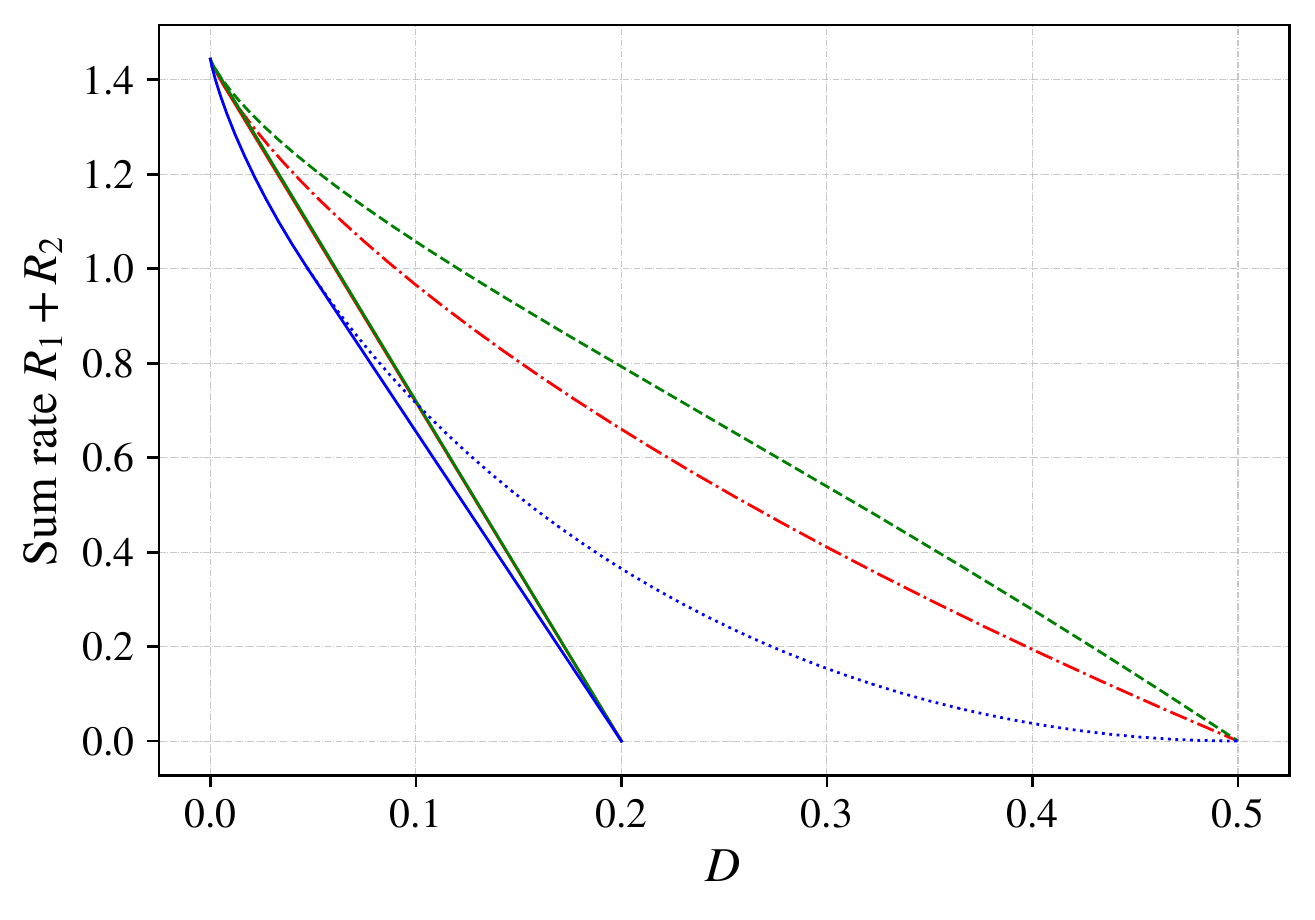}
	\caption{Inner and outer bounds for a DSBS$(p)$ source when $R_1=R_2$ with $p=0.4$ (top) and $p=0.2$ (bottom). 
	}
	\label{fig:inner_outer_bounds}
\end{figure}

The outer bound to $\RR{CSR}$ can be summarized as follows. 
\begin{prop} \label{prop:propositionWZ}
	Let $(X_1, X_2)$ be a DSBS$(p)$ with $0\leq p\leq\frac{1}{2}$. 
	Then, 
	$
	\RR{TSE} = \{ (R_1,R_2,D) \mid  R_1, R_2\geq\Rate{WZ}(D), D\geq0 \},
	$
	and since $\RR{CSR}\subseteq\RR{TSE}$, $\RR{TSE}$ is an outer bound to $\RR{CSR}$. 
\end{prop}
The characterization of $\RR{TSE}$ in Proposition~\ref{prop:propositionWZ} directly follows from the results in \cite{wyner1976rate}. 
In general, it can be shown that the two WZ systems can be used to characterize $\RR{TSE}$ as long as the distortion measure takes the form $\distMeasureForm$ for some function $\combinefunc$. Both Hamming (for binary inputs) and squared distortion measures are of this form. 
In related work, the authors of \cite{yamamoto1982wyner} study WZ theory for a general function of correlated sources. The authors present findings similar to Proposition~\ref{prop:propositionWZ} 
(see therein $X_7$ in Table~1 and Theorem~3). 

\section{Numerical comparison of bounds} \label{secn:comparebounds}

\fig{inner_outer_bounds} illustrates the inner and outer bounds (solid lines) in Propositions 
\ref{prop:propositionSteinberg}, \ref{prop:propositionLKM} and \ref{prop:propositionWZ}, 
for the special case $R_1=R_2$ when $(X_1,X_2)$ is a DSBS$(p)$. 
Let $G(q)=\Hb(p*q)-\Hb(q)$ as defined in \cite{wyner1976rate} and \cite{steinberg2009coding}. 
Non-solid lines in the three cases correspond to $G(q)$ plotted against $q*q$, 
$\Hb(p*q*q)-\Hb(q)$ plotted against $q*q$, and $G(q)$ plotted against $q$, respectively. 
In each case the straight portion of the solid line (i.e., the large $D$ portion) is tangent to the corresponding non-solid line. This results from time-sharing with the zero-sum-rate point. The inner and outer bounds get tighter as $p$ decreases (e.g., the bounds are tighter in right-hand figure compared to the left-hand figure). 

Recall that an optimal scheme for the CSR problem will 
(1) exploit fully the sources observed at each terminal as side information when reconstructing the sum, and 
will (2) exploit the fact that only the sum of the sources needs to be reconstructed. 
Out of the two requirements, only (1) applies to the inner bound obtained with Steinberg's problem, 
and only (2) applies to the inner bound obtained with LKM problem. 
We observe in the figures that the inner bound with Steinberg's problem gives a slightly tighter bound than that with LKM problem (i.e., the red plots are slightly below the green plots).

\section{Conclusion} \label{secn:conclusion}
In this paper we present the problem of two-terminal source coding with Common Sum Reconstruction (CSR), 
in which two terminals want to reconstruct separately the sum of two correlated sources under the constraint that the two reconstructions must be identical with high probability. 
In practice, the CSR problem setup can be observed in algorithms such as butterfly all-reduce \cite{thakur2005optimization}. 
We develop two inner and outer bounds to the achievable rate distortion region of the CSR problem for a doubly symmetric binary source. 
Specifically, we develop two inner bounds using the achievability results for Steinberg's common reconstruction and lossy version of \Korner-Marton's modulo-two sum computation problems. 
We develop an outer bound using Wyner-Ziv's source coding with side information problem.

\section*{Acknowledgment}
The authors would like to thank Maxim Goukhshtein of University of Toronto, and Jason Lam and Zhenhua Hu of Huawei Technologies Canada for technical discussions.

\bibliographystyle{unsrt}
\bibliography{main}

\begin{thebibliography}{10}

\bibitem{suresh2017distributed}
Ananda~Theertha Suresh, X~Yu Felix, Sanjiv Kumar, and H~Brendan McMahan.
\newblock Distributed mean estimation with limited communication.
\newblock In {\em Int. Conf. Machine Learning}, pages 3329--3337, Sydney, Aug
  2017.

\bibitem{thakur2005optimization}
Rajeev Thakur, Rolf Rabenseifner, and William Gropp.
\newblock Optimization of collective communication operations in {MPICH}.
\newblock {\em Int. J. of High Performance Computing Applications},
  19(1):49--66, 2005.

\bibitem{steinberg2009coding}
Yossef Steinberg.
\newblock Coding and common reconstruction.
\newblock {\em IEEE Trans.\ Inform.\ Theory}, 55(11):4995--5010, 2009.

\bibitem{korner1979encode}
Janos K{\"o}rner and Katalin Marton.
\newblock How to encode the modulo-two sum of binary sources.
\newblock {\em IEEE Trans.\ Inform.\ Theory}, 25(2):219--221, 1979.

\bibitem{wyner1976rate}
Aaron Wyner and Jacob Ziv.
\newblock The rate-distortion function for source coding with side information
  at the decoder.
\newblock {\em IEEE Trans.\ Inform.\ Theory}, 22(1):1--10, 1976.

\bibitem{lapidoth2014constrained}
Amos Lapidoth, Andreas Mal{\"a}r, and Michele Wigger.
\newblock Constrained source-coding with side information.
\newblock {\em IEEE Trans.\ Inform.\ Theory}, 60(6):3218--3237, 2014.

\bibitem{lu2016binary}
Jian Lu, Yinfei Xu, Ping Zhang, Man Feng, and Qiao Wang.
\newblock Binary lossy coding problem with encoder side information and common
  reconstruction constraint.
\newblock In {\em Int. Conf. Wireless Communications \& Signal Processing},
  pages 1--5. IEEE, 2016.

\bibitem{sefidgaran2015korner}
Milad Sefidgaran, Amin Gohari, and Mohammad~Reza Aref.
\newblock On {K}{\"o}rner-{M}arton's sum modulo two problem.
\newblock In {\em Iran Workshop on Communication and Information Theory}, pages
  1--6. IEEE, 2015.

\bibitem{nair2020optimal}
Chandra Nair and Yan~Nan Wang.
\newblock On optimal weighted-sum rates for the modulo sum problem.
\newblock In {\em Proc.\ Int.\ Symp.\ Inform.\ Theory}, Los Angeles, Jun 2020.

\bibitem{krithivasan2009lattices}
Dinesh Krithivasan and S.~Sandeep Pradhan.
\newblock Lattices for distributed source coding: Jointly gaussian sources and
  reconstruction of a linear function.
\newblock {\em IEEE Trans.\ Inform.\ Theory}, 55(12):5628--5651, 2009.

\bibitem{wagner2010distributed}
Aaron~B Wagner.
\newblock On distributed compression of linear functions.
\newblock {\em IEEE Trans.\ Inform.\ Theory}, 57(1):79--94, 2010.

\bibitem{yang2014distributed}
Yang Yang and Zixiang Xiong.
\newblock Distributed compression of linear functions: Partial sum-rate
  tightness and gap to optimal sum-rate.
\newblock {\em IEEE Trans.\ Inform.\ Theory}, 60(5):2835--2855, 2014.

\bibitem{lim2019towards}
Sung~Hoon Lim, Chen Feng, Adriano Pastore, Bobak Nazer, and Michael Gastpar.
\newblock Towards an algebraic network information theory: Distributed lossy
  computation of linear functions.
\newblock In {\em Proc.\ Int.\ Symp.\ Inform.\ Theory}, pages 1827--1831. IEEE,
  2019.

\bibitem{goblick1963coding}
T.~J. Goblick.
\newblock {\em Coding for a discrete information source with a distortion
  measure}.
\newblock PhD thesis, Massachusetts Institute of Technology, 1963.

\bibitem{berger1975advances}
T.~Berger and L.~D. Davisson.
\newblock {\em Advances in source coding}.
\newblock Springer, 1975.

\bibitem{yamamoto1982wyner}
Hirosuke Yamamoto.
\newblock Wyner-ziv theory for a general function of the correlated sources.
\newblock {\em IEEE Trans.\ Inform.\ Theory}, 28(5):803--807, 1982.

\end{thebibliography}
\end{document}